\documentclass[journal, lettersize]{IEEEtran}
\usepackage{amsmath,amsfonts}
\usepackage{algorithmic}
\usepackage{algorithm}
\usepackage{array}
\usepackage[caption=false,font=normalsize,labelfont=sf,textfont=sf]{subfig}
\usepackage{textcomp}
\usepackage{stfloats}
\usepackage{url}
\usepackage{verbatim}
\usepackage{graphicx}
\usepackage{cite}
\usepackage{setspace}
\usepackage{svg}
\usepackage{xcolor,colortbl}
\definecolor{lavender}{rgb}{0.9, 0.9, 0.98}

\begin{document}

\title{Open Set Wireless Signal Classification: Augmenting Deep Learning with Expert Feature Classifiers}

\author{Samuel R. Shebert, ~\IEEEmembership{Student Member,~IEEE}, Benjamin H. Kirk, ~\IEEEmembership{Member,~IEEE}, \\ and R. Michael Buehrer, ~\IEEEmembership{Fellow,~IEEE}
\thanks{The support of the US Army Research Office (ARO) is gratefully acknowledged. 
The authors acknowledge Advanced Research Computing at Virginia Tech for providing computational resources that have contributed to the results reported within this paper, URL: https://arc.vt.edu/.
Portions of this paper were presented at GLOBECOM 2021 \cite{globecomConfPaper}, MILCOM 2021 \cite{milcomCofPaper}, and MILCOM 2022 \cite{milcomPaper2}.}}

\maketitle
\begin{abstract}
In shared spectrum with multiple radio access technologies, wireless standard classification is vital for applications such as dynamic spectrum access (DSA) and wideband spectrum monitoring. 
However, interfering signals and the presence of unknown classes of signals can diminish classification accuracy. 
To reduce interference, signals can be isolated in time, frequency, and space, but the isolation process adds distortion that reduces the accuracy of deep learning classifiers. 
We find that the distortion can be partially mitigated by augmenting the classifier training data with the signal isolation steps. 
To address unknown signals, we propose an open set hybrid classifier, which combines deep learning and expert feature classifiers to leverage the reliability and explainability of expert feature classifiers and the lower computational complexity of deep learning classifiers. 
The hybrid classifier reduces the computational complexity by 2 to 7 times on average compared to the expert feature classifiers, while achieving an accuracy of 95\% at 15 dB SNR for known signal classes. The hybrid classifier manages to detect unknown classes at nearly 100\% accuracy, due to the robustness of the expert feature classifiers.
\end{abstract}
\section{Introduction}
\IEEEPARstart{T}{he} rapid growth of radio-based communication systems has triggered interest in using  dynamic spectrum access (DSA) to efficiently access underutilized frequency bands \cite{introSpectrum}. DSA-enabled cognitive radios must often share spectrum with incumbent legacy systems that cannot dynamically avoid interference. \cite{Tripathi2014}. Therefore, cognitive radios must ensure that they do not interfere with transmissions from incumbent users. A key part of avoiding interference with incumbent users is distinguishing the transmissions of incumbent users from secondary users. One way of achieving this is to equip cognitive radios with signal classification algorithms that classify incoming transmissions. Then, the information provided by signal classification could be used to update the spectrum access policy of the cognitive radio and avoid incumbent users.

Signal classification algorithms can be divided into likelihood-based (LB) and feature-based (FB) approaches \cite{Li2019}. LB approaches can achieve theoretically optimal classification accuracies, but in practice are difficult to model for complex classification problems and have high computational complexity. In previous works, maximum likelihood classifiers were derived for modulation \cite{likeModClass,likeRatioModClass}. However, to the best of our knowledge, LB classifiers for standard classification would be too complex for practical use. On the other hand, FB approaches take advantage of distinctive features present in wireless signal standards and are more tractable than LB approaches. 
Therefore, FB approaches are frequently chosen for wireless signal identification.

FB approaches can further be narrowed down to feature design and feature learning. Feature design approaches use features selected by an expert to distinguish signals. Previous works have used expert features such as center frequency, bandwidth, or cyclostationary properties of the signals. In \cite{FuzzyLogic}, the bandwidth and frequency hopping characteristics of signals are used with a fuzzy logic classifier to recognize signal standards. The approach in \cite{babar2013} uses bandwidth and signal pilots to classify GSM, UMTS, and WiFi signals. The authors in \cite{Weng2014} use the bandwidth and periodicity of signals in the ISM band to identify WiFi, Bluetooth, and microwave oven interference. In \cite{alfi2019}, detection of generalized frequency division multiplexing (GFDM) signals was accomplished using the cyclostationary properties of the signals. The authors in \cite{8108218} use a combination of bandwidth, synchronization signals specific to certain standards, and modulation index to identify signals in the ISM band. 

For some wireless signals, it can be difficult to identify clear features that differentiate them. This has led to the use of feature learning approaches which learn relevant features from training examples with limited expert feature design \cite{Li2019}. The feature learning capabilities, combined with high classification accuracy, have made deep learning approaches a popular option for signal classification.

\IEEEpubidadjcol

Classification of wireless signals using deep learning can be segmented into a few primary applications: automatic modulation classification (AMC), RF fingerprinting, and wireless standard identification (WSI). The goal of AMC is to classify between different modulation types, such as AM, FM, PSK, or QAM. Convolutional Neural Networks (CNN) have been proven to be effective at this task \cite{oshea2018, Hermawan2020, Liu2017}. The purpose of RF fingerprinting is to identify the specific emitter of a wireless signal. Deep learning algorithms have been used to extract features from wireless signals that a particular emitter imparts on its transmissions \cite{rfFingerprintDeep, rfFingerprintResnet, zhang2016}. 

The goal of WSI is to identify the wireless standard, such as 4G LTE or IEEE 802.11, of an unknown signal. This differs from AMC because a wireless standard often includes many modulation types, which can be dynamically changed based on channel quality, transmit scheme, or other factors. Table \ref{tab:classifier_review} gives an overview of previous works in the area of deep learning-based WSI. In the table, synthetic refers to simulated signals with modeled channel impairments and OTA and OTW refer to over-the-air and over-the-wire signals, respectively. 
Deep neural networks (DNN), CNN, and long short-term memory (LSTM) recurrent neural networks (RNN) are commonly used deep learning architectures.

\begin{table}[t]
    \caption{Previous Deep Learning-Based WSI Works}
    \label{tab:classifier_review}
    \footnotesize
    \begin{tabular}{| p{\dimexpr 0.14\linewidth-2\tabcolsep} p{\dimexpr 0.14\linewidth-2\tabcolsep} p{\dimexpr 0.44\linewidth-2\tabcolsep} p{\dimexpr 0.28\linewidth-2\tabcolsep} |}
        \hline
        Author & Model & Standards & Signal Generation\\
        \hline
        \hline
        \cite{Alshathri2019} & DNN & GSM, UMTS, LTE & Synthetic\\
        \hline
        \cite{Alhazmi2020} & CNN & UMTS, LTE, 5G & Synthetic\\
        \hline
        \cite{Xia2019} & CNN & GSM, UMTS, LTE & Synthetic\\
        \hline
        \cite{hiremath2018} & CNN & GSM, Bluetooth, WiFi & Synthetic\\
        \hline
        \cite{Ahmed2017} & CNN & WLAN, LTE, Radar & Synthetic\\
        \hline
        \cite{Naim2017} & CNN & WiFi, Bluetooth, Zigbee & Synthetic\\
        \hline
        \cite{Merima2017} & CNN & WiFi, Bluetooth, Zigbee & Synthetic\\
        \hline
        \cite{Behura2020} & CNN & UMTS, LTE, WiFi, Bluetooth, +24 more & Synthetic/OTA\\
        \hline
        \cite{zhang2019} & CNN/ LSTM & WiFi, Bluetooth, Zigbee & OTW\\
        \hline
        \cite{schmidt2017} & CNN & WiFi, Bluetooth, Zigbee & OTW\\
        \hline
    \end{tabular}
\end{table}

Most of the previous works in Table \ref{tab:classifier_review} do not look at the  full scope of a spectrum monitoring technology. With the exception of interference detection classifiers \cite{zhang2019, schmidt2017}, the deep learning architectures used require the signal of interest to be isolated from all other signals in the spectrum to get an accurate classification. This can be particularly difficult if devices are interfering in time and frequency with one another. One solution is to isolate signals in time, frequency, and space using signal processing techniques prior to classification. However, signal isolation requires estimation of signal parameters, such as the number of signals interfering in time and frequency, the center frequency and bandwidth, and duration, which propagates estimation error to the classifier. In the case of deep learning classifiers, the loss in accuracy due to the estimation errors can be partially mitigated through data augmentation on the training dataset.

In addition, the previous works in Table \ref{tab:classifier_review} are \textit{closed set}, meaning they must classify signals as one of the standards included in the training set. The drawback to closed set classification is the inability to appropriately classify unknown signal standards that the classifier was not trained to recognize. In practice, the large amount of existing signal standards, combined with the perpetual development of new standards makes it difficult to include all signal classes in the training data. Therefore, a wireless standard classifier would need to identify signals it was not pretrained to classify to appropriately describe a signal environment.

\begin{figure}
    \centering
    \includegraphics[scale=0.20]{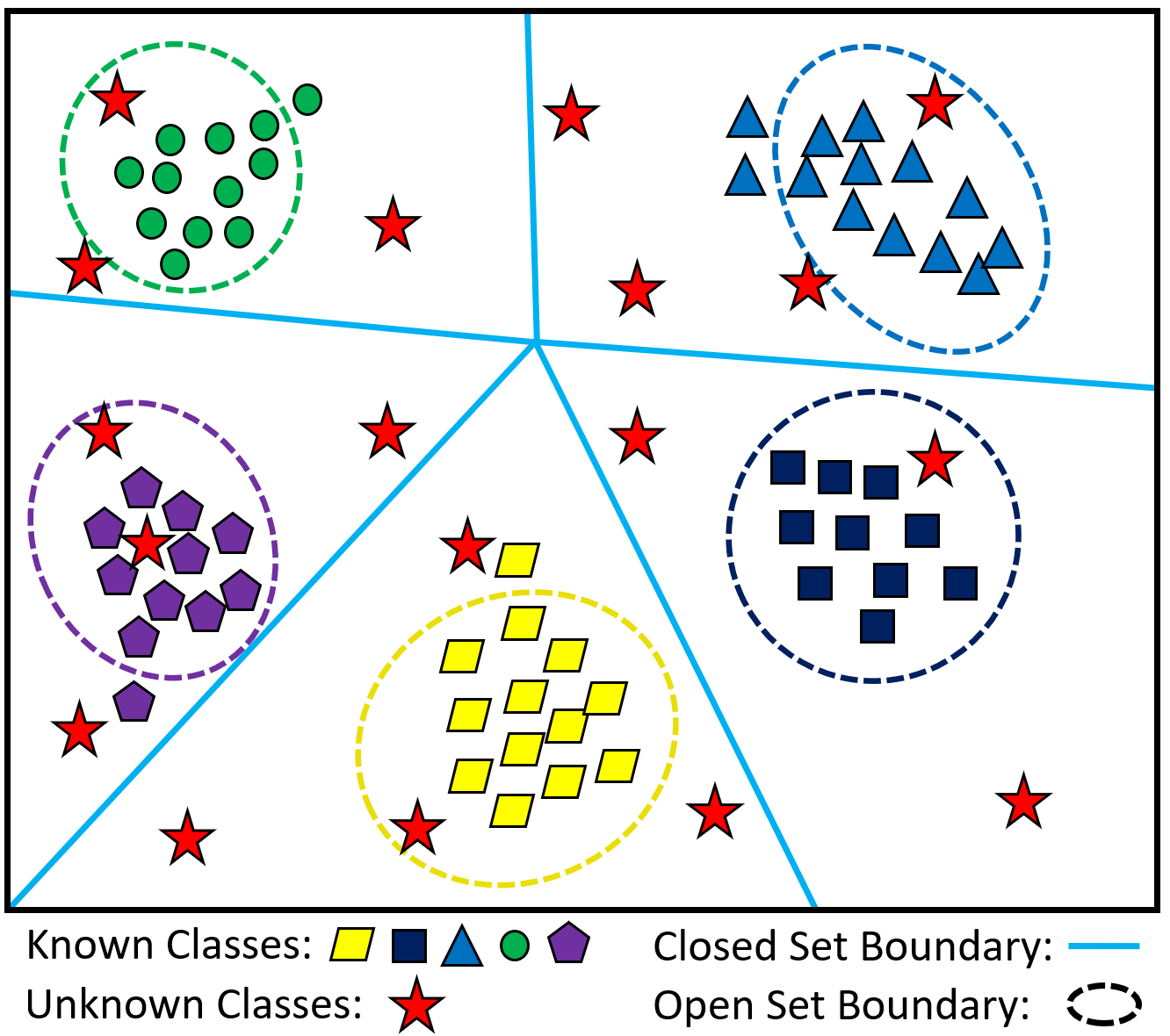}
    \caption{Example distribution of signals from known and unknown classes in an arbitrary feature space. Solid lines indicate the decision boundaries of a closed set classifier. Dashed lines indicate shrunken decision regions used by open set classifiers.}
    \label{fig:openSet}
\end{figure}

This drawback is addressed through the development of ``open set" classifiers.
The goal of an open set classifier is to detect data from ``unknown" classes, while still accurately classifying data from the ``known" classes \cite{Geng2020}. 
In the context of deep learning classifiers, data from ``known" classes are included in the training dataset and data from classes that are not represented in the training class are considered "unknown".
The concept of open set classification is illustrated in Figure \ref{fig:openSet}.
Five clusters of known classes are distributed in an arbitrary feature space.
A closed set deep learning classifier learns decision boundaries that minimize the loss incurred for the training data. 
This gives the best accuracy when the classifier only needs to classify known classes. 
However, if data from unknown classes are distributed throughout the feature space, they will be incorrectly classified as the class of whichever decision region they fall into. 
The aim of open set classification is to tighten decision boundaries around the known classes, such that data from unknown classes do not fall into any decision region. 
Then, input data that does not fall into any known class can be identified as unknown. 
In addition to correct classifications, Figure \ref{fig:openSet} depicts two undesirable outcomes: false negatives and false positives.
False negatives occur when known signals are identified as unknown because they fall outside of the open set boundaries and false positives occur when unknown signals are inside of the open set boundaries.  

Bendale and Boult developed an open set detector in the form of a new layer for deep learning models, called OpenMax \cite{Bendale2016}. This layer acts as unknown class detector for open-set images or adversarially generated images. The authors of \cite{Shu2017} proposed to threshold the sigmoid activated output values of a neural network to reject unknown signal classes and found their method outperformed OpenMax for text classification. In the field of signal classification, Gong et. al created an open set deep learning classifier to identify signals in the ISM band \cite{ismOSR2022}. However, a drawback of purely deep learning-based open set classifiers is the inability to identify unknown classes that are similar to known classes. When this is coupled with the lack of explainability of deep learning models, it can be challenging to trust deep learning classifications. We show that a hybrid approach, using both deep learning and expert feature classifiers, can add explainability and robustness to open set classification.

In this paper, we propose an open set CNN-based wireless standard classifier that uses a modified version of the unknown class detector originally developed for text classification by the authors in \cite{Shu2017}.
The CNN classifier is trained with 4G LTE downlink and uplink, 5G NR downlink and uplink, IEEE 802.11ax (WiFi 6), Bluetooth Low Energy (BLE) 5.0, and Narrowband Internet-of-Things (NB-IoT) signals generated using MATLAB. 
The training signals are augmented with Rayleigh or Rician fading, AWGN, frequency offsets, spurious or DC components, and in-phase/quadrature (I/Q) imbalances to model the wireless channel and radio hardware imperfections.
Additionally, we augment training signals with signal detection and isolation steps. 
This augmentation allows the CNN to mitigate estimation errors and distortion caused by necessary steps for wideband spectrum sensing.
The CNN classifier is compared against a bank of expert feature classifiers developed for each standard. 
The expert feature classifiers are based on the synchronization algorithms that are typically used for the initial connection of devices.
The open set capabilities are tested on a set of unknown signal classes made up of generic orthogonal frequency-division multiplexing (OFDM), single carrier frequency-division multiple access (SC-FDMA), single carrier (SC), amplitude modulation (AM), and frequency modulation (FM) signals. 
This extends previously published conference papers \cite{globecomConfPaper, milcomCofPaper, milcomPaper2} by introducing expert feature classifiers and considering signal detection and isolation. 
Specifically, the deep learning feature preprocessing method from \cite{globecomConfPaper}, the open set unknown class detector from \cite{milcomCofPaper}, and the co-frequency signal classification system from \cite{milcomPaper2} were used in this paper. The new contributions are as follows:
\subsubsection{Impact of Signal Detection and Isolation on Wireless Standards Classification}
This paper shows that automatic signal detection and isolation necessary for wideband spectrum sensing can negatively impact signal classification.
In particular, center frequency and bandwidth estimation and frequency domain filtering add estimation error and distortion that reduce classification accuracy.
For deep learning classifiers, we show that including automatic frequency isolation in the data augmentation steps for the training dataset offsets some of the losses in accuracy.

\subsubsection{Comparison of Deep Learning and Expert Feature Classifiers}
Deep learning algorithms are a prevailing option for solving challenging classification problems. However, wireless standards are designed with features that allow devices to detect and estimate parameters of the signals. This paper compares the performance of the deep learning and expert feature classifiers in terms of accuracy and computational complexity. Then, we propose a hybrid classifier that obtains the benefits of both approaches.

\begin{figure*}
    \centering
    \includegraphics[scale=0.45]{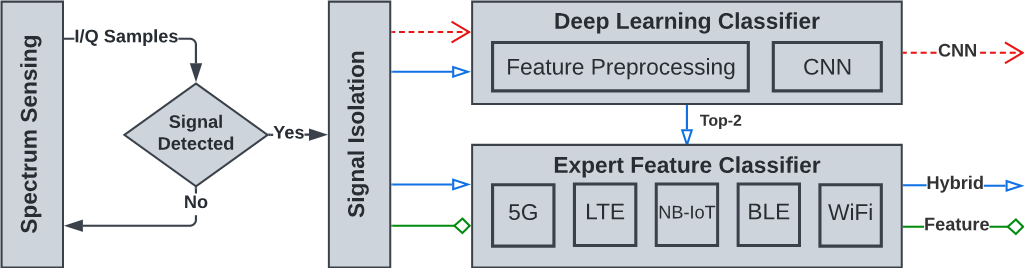}
    \caption{Flowchart of CNN, expert feature, and hybrid classifier. After the ``Signal Isolation" block, the data paths of each classifiers diverges, denoted by the changes in arrow color and symbol.}
    \label{fig:systemFlowChart}
\end{figure*}

\subsubsection{Open Set Wireless Standard Classification}
In uncontrolled spectral environments, it is difficult to design a closed set classifier to classify any type of wireless signal that could be present. This indicates that signal classification systems must be able to identify signals from unknown classes to appropriately categorize signals in a spectral environment. 
In this paper, we apply a discriminative unknown class detector to a deep learning classifier for wireless standard classification. Additionally, we show that expert feature classifiers can be used to augment the robustness of the open set deep learning-based classifier while maintaining a computational complexity advantage over pure feature-based approaches.

\section{System Model}
The proposed classifier system is shown in Figure \ref{fig:systemFlowChart}. The ``Spectrum Sensing" block encompasses the RF front-end processes, including amplification, filtering, downconversion, and analog to digital conversion that produce I/Q samples. The following blocks are discussed in the next sections.

\subsection{Signal Detection}
Assuming no prior knowledge about the signal is known, energy detection is the best option for signal detection \cite{sigDetection2011}. Deep learning algorithms have been proposed as an alternative for signal detection and time-frequency localization \cite{osheaDetection}; however, such an approach is outside the scope of this work. Classical energy detection can be described as a hypothesis test:
\begin{equation}
    \begin{split}
    \mathcal{H}_0 &: r(t) = n(t)\\
    \mathcal{H}_1 &: r(t) = hx(t) + n(t)
    \end{split}
\end{equation}
$\mathcal{H}_0$ is the null hypothesis where no signal is present and $\mathcal{H}_1$ is the case where a signal is present. $r(t)$ is the received signal, $n(t)$ is additive white Gaussian noise (AWGN), $h$ is a complex channel gain, and $x(t)$ is the signal of interest \cite{sigDetection2011}.
A computationally efficient energy detection algorithm for fast wideband spectrum sensing developed by Martone and Ranney \cite{fss2014} is implemented for both signal detection and time-frequency localization. The algorithm uses an energy threshold to detect signals in the frequency domain, then iteratively groups FFT frequency bins to estimate the number of signals and the bandwidth of each signal. 

\subsection{Signal Isolation}
Signal isolation refers to techniques that can separate mixtures of signals. 
In this paper, we assume that interfering signals have a form of orthogonality, either in time, frequency, or space, that can be exploited to filter the received signals. 

\subsubsection{Time}
When the signals of interest are orthogonal in time, isolation simply requires selecting the I/Q samples corresponding to the transmission time of each signal. The time span of each signal is estimated by splitting a long capture of I/Q samples into short 1 ms segments and performing energy detection on each segment independently.
\subsubsection{Frequency} 
In this case, the signals were transmitted at different carrier frequencies with non-overlapping bandwidths.
The center frequency and bandwidth of the signals are estimated using the energy detection algorithm in \cite{fss2014}. Then, the signal is digitally frequency-shifted to center in complex baseband. If the signal will be classified by the CNN, a 5$^{th}$ order Butterworth lowpass filter is used with a cutoff frequency equal to the estimated bandwidth. In the case of the expert feature classifiers, a Kaiser windowed FIR lowpass filter is used with cutoff frequency and order depending on the standard. These lowpass filter designs were primarily chosen for convenience and other well designed filters would likely perform similarly.
\subsubsection{Space}
If the signals are transmitted from different directions spatially, they will impart different signatures across an array of receive antennas. A uniform linear array (ULA) is used for simplicity, which has the following $M_r \times 1$ signal vector, $\mathbf{a}(\theta)$:

\begin{equation}
    \mathbf{a}(\theta) = \begin{bmatrix} 1 \\ e^{-j\frac{2\pi d}{\lambda} sin(\theta)} \\ e^{-j2\frac{2\pi d}{\lambda} sin(\theta)} \\ \hdots \\ e^{-j(M_r-1)\frac{2\pi d}{\lambda} sin(\theta)}\end{bmatrix}
\end{equation}
where $M_r$ is the number of antennas, $\theta$ is the angle of arrival relative to the boresight of the array, $d$ is the distance between array elements, and $\lambda$ is the wavelength. If $x_i[k]$ is transmitted from angle $\theta_i$, the received signal will be:

\begin{equation}
    \mathbf{r}[k] = \sum_{i=1}^M \mathbf{a}(\theta_i)x_i[k] + \mathbf{n}[k]
\end{equation}
where $\mathbf{n}[k]$ is independent AWGN with $\Sigma_n=\sigma^2 I$ and $M$ is the number of signals. An estimate of the covariance of the received signal is formed as:
\begin{equation}
    \hat{\mathbf{R}} = \frac{1}{N} \mathbf{r} \mathbf{r}^H
\end{equation}
where $(.)^H$ is the Hermitian conjugate transpose and $N$ is the number of samples in $\mathbf{r}$. For $M$ emitters, the eigenvectors of $\hat{\mathbf{R}}$ corresponding to the $M$ largest eigenvalues define the signal subspace. The remaining $M_r - M$ eigenvalues correspond to the noise sub-space \cite{moe1991}. Therefore, the approach detailed in Algorithm \ref{alg:moe} uses a threshold on the eignvalues to estimate $M$. This method is able to detect up to $M_r - 1$ signals that could be interfering in time and frequency. The threshold, $\tau$, controls probability of detection $P_d$ and false alarm $P_{fa}$ of the estimator.
The value of $\tau$ was chosen conservatively such that the $P_{fa}$ is well below 1\%.
\begin{algorithm}
    \caption{Model Order Estimation}
    \label{alg:moe}
    \begin{algorithmic}
        \STATE $\hat{M} \gets 0$
        \STATE $\tau \gets 1.1$
        \STATE $\boldsymbol{\lambda} \gets \text{eigen}(\hat{\mathbf{R}})$
        \STATE $\lambda_{noise} \gets \text{min}(\boldsymbol{\lambda})$
        \FOR{$i=1,\dots,M_r-1$}
            \IF{$\lambda_i > \tau \cdot \lambda_{noise}$}
                \STATE $\hat{M} \gets \hat{M} + 1$
            \ENDIF
        \ENDFOR
    \end{algorithmic}
\end{algorithm}

After estimating $M$, a direction of arrival (DOA) estimation algorithm, such as MUSIC \cite{music1986}, can be used to estimate the angles of each signal:
\begin{equation}
    \hat{\mathbf{\theta}} = \text{MUSIC}(\hat{\mathbf{R}}, \hat{M})
\end{equation}
Next, the spatial filter weights for each received signal are created using the minimum mean square error (MMSE) beamformer:
\begin{equation}
    \mathbf{w}_i = \hat{\mathbf{R}}^{-1} \mathbf{a}(\hat{\theta_i})
\end{equation}
Finally, each co-frequency signal can be recovered by filtering the received signal:
\begin{equation}
\hat{x}_i[k] = \mathbf{w}^H_i \mathbf{r}[k]
\end{equation}

\subsection{Feature Preprocessing for Deep Learning} Feature preprocessing is limited expert feature design used to improve the accuracy of deep learning models. Although deep learning models can learn features on their own, simple feature preprocessing steps often help models learn better solutions.

\subsubsection{Signal Domain Transformation}
Signals can be represented in multiple domains, including time, frequency, or joint time-frequency domains. In this work, signals are converted to a time-frequency representation using multiple Fast Fourier Transforms (FFT). 131,072 received I/Q samples (corresponding to approximately 1 ms duration at a sampling rate of 125 MHz) are split into two, 65,536 sample segments. Then, the FFT of each segment produces a high frequency resolution time-frequency signal representation. This feature was determined experimentally to be advantageous for CNN learning in \cite{globecomConfPaper}.

\subsubsection{Magnitude}
In this step, the magnitude of the complex valued samples is taken to reduce the memory footprint and the computational complexity in subsequent processing. This results in a loss of the phase information in the signal, but it was determined experimentally that this had a negligible impact on wireless standard classification.

\subsubsection{Normalization}
The final preprocessing step scales signals between 0 and 1 to ensures that the model does not use the signal power as a feature.

\begin{figure}[t]
    \centering
        \includegraphics[width=0.78\linewidth]{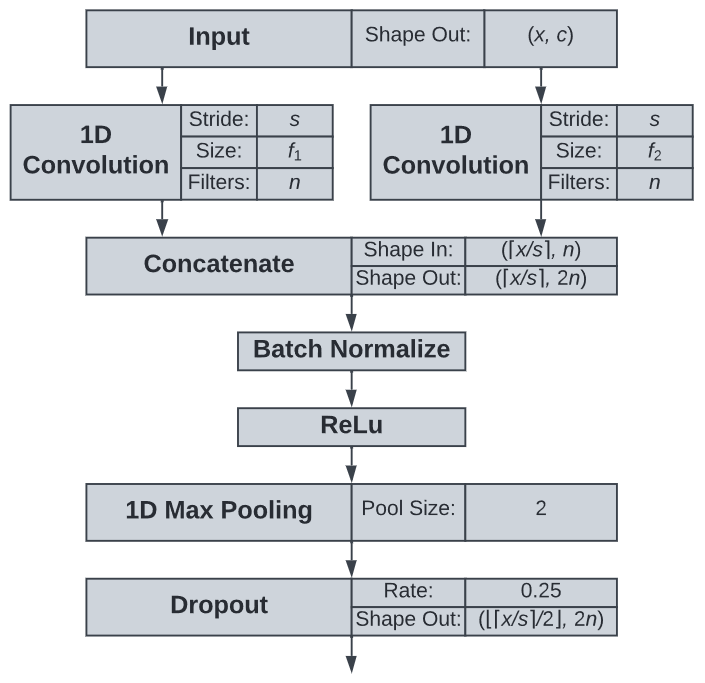}
        \caption{Structure of the "Convolution Block". The block is parameterized by the stride, $s$, the number of filters, $n$, and the filters sizes, $f_1$ and $f_2$. The input to the block is $c$, 1 dimensional vectors of length $x$.}
        \label{fig:convBlock}
\end{figure}

\subsection{CNN Architecture}
The CNN architecture designed for this work is inspired from the inception modules created by Szegedy et al. \cite{inceptionv1}. The primary block in the architecture, refered to as a ``Convolution Block", is shown in Figure \ref{fig:convBlock}. The advantage of this block is that the different filter sizes in the 1D convolutions allow the model to learn a more diverse set of features than standard CNNs. The outputs of the two convolutions are concatenated, which results in an output with twice the number of filters. Batch normalization is used to improve the convergence of the model by correcting internal covariance shifts \cite{batchNorm}. This is followed by the ReLU activation function and a 1D max pooling that decimates by a factor of 2. Finally, a dropout layer with a rate of 0.25 helps regularize the model \cite{dropout}.

\begin{figure}[t]
        \centering
        \includegraphics[width=\linewidth]{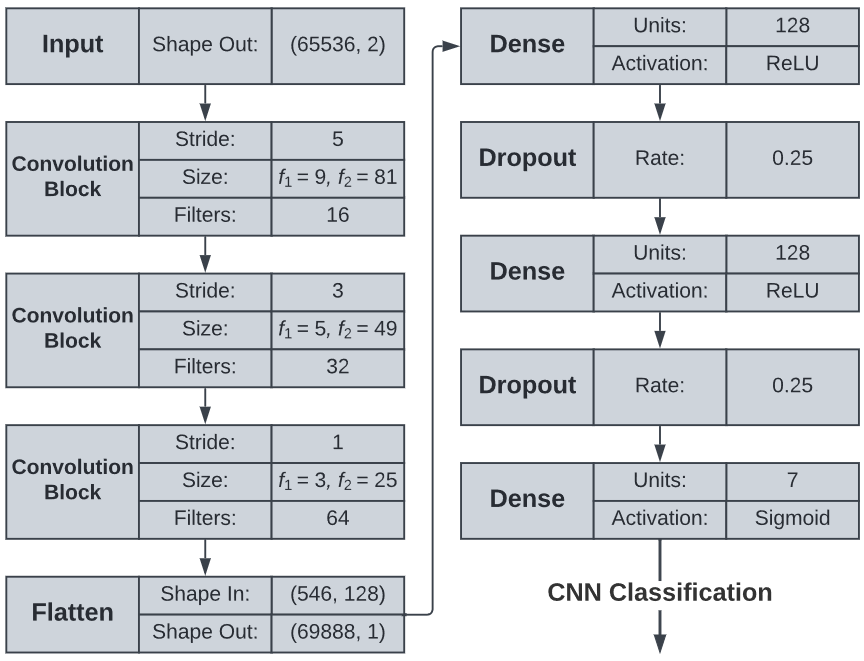}
        \caption{Full CNN architecture. The input is two, 65536 point DFTs. The output is a vector with 7 elements. Each element corresponds to the models estimated probability of the input signal being that class. "Convolution Blocks" are defined in Figure \ref{fig:convBlock}}
        \label{fig:mlModel}
\end{figure}

The full model, shown in Figure \ref{fig:mlModel}, consists of three convolution blocks followed by dense and dropout layers. The model weights are initialized using the Glorot uniform initializer \cite{glorot} and the model is trained with a batch size of 300 and a learning rate of 0.001.

\subsection{Unknown Class Detector for Deep Learning}
The unknown class detector thresholds the output of the CNN to identify signals that do not belong to any known class. The effectiveness of a thresholding approach partially depends on the activation function used at the output of the classifier model. The softmax activation function is commonly used for multi-class classification, but is not suitable for use in open set problems. The softmax function for the $i$th class of an $N$ class classifier is as follows:
\begin{equation}
    \text{softmax} (y_i) = \frac{e^{y_i}}{\sum^N_{j=1}e^{y_j}}
\end{equation}
Where $y_i$ is the $i$th value of the $N\times1$ output vector from the classifier. The term in the denominator normalizes the output of the softmax function, such that the sum of all of the softmax values will be 1. However, normalization is only reasonable if the classes are collectively exhaustive, such that the probability of all events occuring must sum to 1. This assumption is broken in open set classification, which is premised on the existence of additional classes that are not in the training set.
On the other hand, the sigmoid function does not assume that the classes are collectively exhaustive. The sigmoid function for the $i$th class is as follows:
\begin{equation}
    \text{sigmoid}(y_i) = \frac{1}{1+e^{-y_i}}
\end{equation}
The sigmoid function independently estimates the probability of each class, which is important for open set classification \cite{Shu2017}.

The unknown class detector thresholds the sigmoid output values, with values below the threshold considered to be unknown and values above the threshold to be known.
The unknown class detection algorithm for a classifier with \textit{N} classes is shown in Algorithm \ref{alg:openset}. If any of the output values from the CNN are greater than the sigmoid function of the threshold, $\eta$, then the predicted signal class, $\hat{y}$, is the index of the greatest CNN output value. Otherwise, $\hat{y}$ is unknown and is assigned a value of $N+1$.

\begin{algorithm}
    \caption{Unknown Class Detector}
    \label{alg:openset}
    \begin{algorithmic}
        \STATE $N \gets \text{ number of known classes}$
        \STATE $\mathbf{y} \gets N\times1\text{ CNN output vector}$
        \STATE $\eta \gets \text{ detector threshold} $
        \IF{$\text{any }\mathbf{y} > \text{sigmoid(}\eta\text{)} $}
            \STATE $\hat{y} \gets \text{argmax(}\mathbf{y}\text{)}$
        \ELSE
            \STATE $\hat{y} \gets N+1$
        \ENDIF
    \end{algorithmic}
\end{algorithm}

\subsection{Expert Feature Classifiers}
Wireless standards often contain robust features for detection in the form of synchronization signals, which can be used to identify signals from that standard.
Synchronization signals are included in transmissions so that the receiver can estimate the timing and frequency offsets of a received signal. Many wireless standards use a fixed set of synchronization signals that are known \textit{a priori} by the receiver. 
Once the timing and frequency offset have been corrected, the received signal can be demodulated according to the standard and the contents of the message can be verified. 
This is done by decoding a Cylic Redundancy Check (CRC) code appended to the message by the transmitter.
A requirement for CRC decoding is that the transmitter and receiver agree on the same generator polynomial \cite{crc}. 
In most cases, the generator polynomial is defined \textit{a priori} in the standard. 
A bank of synchronization-based classifiers, made up of 4G and 5G downlink, WiFi 6, BLE 5, and NB-IoT form a classifier that can classify the same standards as the CNN classifier. Each classifier makes a binary decision indicating whether the signal corresponds to that standard. If the data integrity check of a classifier passes, then the classifier decides that the corresponding standard has been detected. Compared to the proposed CNN, these classifiers require a long duration signal sample and are very computationally complex, as we will discuss later. The next sections briefly describe the the synchronization processes used to classify each standards.

\subsubsection{4G LTE Downlink} 
First, each of the three possible primary synchronization sequences (PSS) \cite{lteDLSync} are cross-correlated with the received signal to estimate the timing offset.
This same process can be repeated with frequency shifted copies of the received signal to estimate coarse frequency shifts.
Then, a fine frequency offset estimation algorithm using the cyclic prefix in the OFDM symbols can corrects for smaller frequency shifts between $\pm$7.5 kHz.
After timing and frequency synchronization, the 168 secondary synchronization sequences (SSS) \cite{lteDLSync} are cross-correlated with the extracted and demodulated sequence from the received signal. 
The highest correlation PSS and SSS sequences are used to determine the cell ID, which is required to descramble the master information block (MIB).
An LTE signal is considered to be detected if the 16 bit CRC appended to the MIB message is valid.
The periodicity of MIB transmision, and therefore the minimum sample duration, is 10 ms.




\subsubsection{NB-IoT} 
With a few exceptions, the NB-IoT synchronization procedure is the same as LTE.
The three PSS and 168 SSS in LTE are replaced by a single narrowband PSS (NPSS) and 504 narrowband SSS (NSSS) \cite{lteDLSync}. The MIB periodicity remains at 10 ms, but the NSSS periodicity is 20 ms, requiring a minimum sample duration of 20 ms.

\subsubsection{5G NR Downlink} 
The 5G synchronization procedure shares the same sequence of steps as LTE and NB-IoT, but many details are altered.
The number of SSS codes are increased to 336, the MIB CRC is increased to 24 bits, and the periodicity of PSS, SSS, and MIB transmission is adjustable within the range of 10-160 ms \cite{nrDLSync}.
Therefore, a minimum sample duration of 160 ms is needed to ensure reception of the synchronization signalling, but for simplicity, we assume that the signalling will be transmitted every 10 ms.

\subsubsection{4G and 5G Uplink}
Classification of 4G and 5G uplink signals is challenging due to a lack of persistent broadcast signaling and a high degree of flexibility.
There are a couple of candidate signals embedded in uplink transmissions that could be used for classification, namely the random access signals and the sounding reference signals.
However, these signals lack a CRC, making it difficult to design a low false alarm rate detector. Instead, determining the corresponding downlink signal and checking the MIB CRC is a more reliable option. 

For frequency division duplex (FDD) systems, the downlink signal frequency most likely at the default duplex frequency. 
If a downlink signal is found at the duplex frequency, the downlink signal can be decoded to verify the frequency of the corresponding uplink band. 
The information for the uplink band frequency is found in system information block (SIB) 2 for LTE \cite{lteSIB} or SIB 1 for 5G \cite{nrSIB}. In the case of a time division duplex (TDD), the LTE or 5G downlink signals could be detected and demodulated to determine that the cell is TDD.

Although verifying the downlink SIB is the most reliable option, for simplicity, we implement an approach which only uses the frequency separation between downlink and uplink channels. 
The frequency of detected signals is cross-referenced with a lookup table of default frequency duplex separation values found in the 3GPP standard \cite{lteDuplex, nrDuplex}. If a signal was detected in an uplink frequency band, the receiver would re-tune its frequency to the corresponding downlink frequency band. Then, the LTE and 5G downlink expert feature classifiers would attempt to classify signals detected in the downlink band. If a downlink signal was found, the original signal would be classified as an uplink signal.

There is a concern that in some cases, the downlink signal corresponding to a candidate uplink signal would not be detectable to the classifier. This would be due to the channel between the user equipment (UE) and the classifier being more favorable than the channel between the basestation and the classifier. However, we assume that this would be uncommon because (1) the basestation must be relatively close to the UE (2) the placement of basestations generally make their channel more favorable than the channel to the UE and (3) basestations generally have higher transmit power than UEs.

\subsubsection{WiFi 6} 
WiFi uses a packet based transmission paradigm, making a sliding window packet detector useful to determine the start of a transmission.
Then, a fine timing offset estimation is achieved by cross-correlating the L-STF and L-LTF packet preamble fields \cite{wifiHTPhy} with the received signal. Next, the L-STF and L-LTF fields are reused for frequency offset estimation. The received signal is corrected based on the timing and frequency offset estimates and the packet preamble is demodulated. Finally, the parity bit in the L-SIG field \cite{wifiHTPhy} and the 3 bit CRC in the HE-SIG-A  field \cite{wifiHEPhy} are verified.

\subsubsection{BLE 5}
Similar to WiFi, BLE uses packet based transmissions. There are two primary packet types of interest: advertisement and data packets. Advertisement packets are transmitted on dedicated frequency channels and are used for setting up initial connections or broadcasting short messages. Data packets are transmitted over the remaining frequency channels and carry the bulk of information after a connection between devices has been established. The CRC generator polynomial is known \textit{a priori} for advertisement packets, but not for data packets as a new generator polynomial is negotiated during the initial connection process \cite{bleCRCGen}. As a result, the proposed BLE classifier can only identify advertisement packets.

First, the BLE classifier estimates timing offset by cross-correlating the an advertisement packet preamble with the received signal. Next, the frequency offset is estimated using an FFT-based approach and corrected.
Finally, the synchronized advertisement packet is demodulated and the 24 bit CRC is validated.

\subsection{Expert Feature Classifier Probability of False Alarm}
When no signal is present or the signal does not conform to the wireless standard of the expert feature classifier, the received message after demodulation is modeled as random bits. This results in the $P_{fa}$ being approximately equal to the probability that the CRC passes, given random bits at the receiver. Given a random message and CRC code, the probability of a CRC not detecting an error is \cite{crc}:
\begin{equation}
    P_{fa} \approx \left( \frac{1}{2} \right)^n
\end{equation}
where $n$ is the number of CRC bits. Table \ref{tab:crcPfa} contains the theoretical $P_{fa}$ for each of the feature classifiers. In some cases, the $P_{fa}$ can be smaller than the CRC error rate due to additional checks before the CRC. For example, the WiFi expert feature classifier must detect a packet using a sliding window packet detector before a CRC is attempted. This can cause the $P_{fa}$ to be substantially lower, especially in the case of continuous signals, where the rising edge of packet does not occur.

\begin{table}[t]
    \begin{center}
    \caption{Theoretical $P_{fa}$ of Feature Classifiers}
    \label{tab:crcPfa}
    \small
    \begin{tabular}{|c| c | c|} 
        \hline
        Classifier & $P_{fa}$\\ [0.5ex] 
        \hline\hline
        4G Downlink/Uplink & $1.53\times10^{-5}$\\ 
        \hline
        NB-IoT Downlink & $1.53\times10^{-5}$\\
        \hline
        5G Downlink/Uplink & $5.96\times10^{-8}$\\
        \hline
        BLE Adv. Packet & $5.96\times10^{-8}$\\
        \hline
        WiFi 6 & $3.12\times10^{-2}$\\
        \hline
    \end{tabular}
    \end{center}
\end{table}

\subsection{Minimum Sample Duration}
The minimum sample duration for classification, and equivalently the number of required I/Q samples, varies depending on the classifier. Table \ref{tab:signalDuration} contains the sample duration's required by each classifier. The proposed CNN classifier requires approximately 1 ms of data, while many of the expert feature classifiers require more. In the case of 4G, 5G, and NB-IoT, the synchronization signals are guaranteed to be transmitted periodically, which determines the minimum sample size required. In the case of 5G, the periodicity is flexible, but we assume the periodicity will be 10 ms for simplicity. For WiFi and BLE packets, the minimum sample duration is less than a millisecond to capture the packet preamble and CRC required for detection. 
However, we choose a longer sample duration of 10 ms to increase the chance that the CRC will be captured in the sample.


\begin{table}[t]
    \begin{center}
    \caption{Signal Sample Duration of Classifiers}
    \label{tab:signalDuration}
    \small
    \begin{tabular}{|c| c | c|} 
        \multicolumn{2}{c}{\textbf{Deep Learning}}\\
        \hline
        Classifier & Minimum Sample Duration (ms)\\
        \hline \hline
        CNN & 1\\
        \hline
        \multicolumn{2}{c}{\textbf{Expert Feature}}\\
        \hline
        Classifier & Minimum Sample Duration (ms)\\
        \hline \hline
        4G Downlink/Uplink & 10\\
        \hline
        NB-IoT Downlink & 20\\
        \hline
        5G Downlink/Uplink & 10-160\\
        \hline
        BLE Adv. Packet & 10 \\
        \hline
        WiFi 6 & 10\\
        \hline
    \end{tabular}
    \end{center}
\end{table}

\section{Wireless Signal Datasets}
Three datasets were developed for training and testing the classifiers.
The ``known signal dataset" is a synthetic dataset of wireless standard signals generated in MATLAB 2021a.
This dataset was split 80-10-10 between training, validation, and testing data.
The ``unknown signal dataset" is a synthetic dataset of generic signals generated in MATLAB 2018b used for testing only.
The final dataset is the ``over-the-air dataset" made up of wireless standard signals and only used for testing.
These datasets are detailed in the following sections.

For all datasets, signals were generated using random transport data and varied signal parameters. An overview of the parameters for each wireless standards can be found in Table \ref{tab:signal_standards}.
All signals were sampled at 125 MHz to satisfy the Nyquist rate for all signal bandwidths of interest.
Additionally, all signals were processed by the signal detection and isolation algorithms.



\subsection{Known Signal Dataset} \label{sec:known_dataset}
This dataset is made up of 6720 signals, 960 per wireless standard. 
The duration of each signal varies from 10 to 20 ms, depending on the the minimum sample duration shown in Table \ref{tab:signalDuration}.
Since the CNN classifier only requires 1 ms of data, each signal is bootstrap sampled into 30 1 ms samples, resulting in 201600 signals total.
The procedure for generating signals from each signal standard is described next.
\subsubsection{4G LTE \& 5G NR}
4G LTE downlink and uplink signals were generated to be Release 8 compliant using the MATLAB LTE Toolbox \cite{matlabLTE}. 5G NR downlink and uplink signals were generated to be Release 15 compliant using the MATLAB 5G Toolbox \cite{matlab5G}. Resource block utilization of the shared channels ranged from 80\% to 100\% and the allocation of resource blocks followed two traffic models. The first traffic model randomly allocated resource blocks throughout the radio frame. The second model replicated bursty traffic by allocating large contiguous blocks of resource blocks sporadically in time.

\subsubsection{IEEE 802.11ax (WiFi 6)}
WiFi 6 signals were generated using the MATLAB WLAN Toolbox \cite{matlabWLAN}. The generated signals followed the single-user packet format. Randomized traffic was simulated by using packets with different transport data lengths and inserting periods of no transmission between packets.

\subsubsection{Bluetooth Low Energy (BLE) 5.0}
BLE signals were generated using the MATLAB Bluetooth Toolbox \cite{matlabBLE}. The simulated BLE signals included frequency hopping advertisers, pairing devices, and ongoing data connections.
However, for tests involving the expert feature classifiers, the BLE signals are limited to advertisers.

\begin{table*}[t]
\begin{center}
    \caption{Overview of Signal Standard Parameters}
    \label{tab:signal_standards}
    \footnotesize
    \begin{tabular}{ |c | c | c | c | c | c | c | c |}
        \hline
        & 4G Downlink & 5G Downlink & 4G Uplink & 5G Uplink & WiFi 6 & BLE 5.0 & NB-IoT\\
        \hline
        \hline
        Waveform & OFDMA & OFDMA & SC-FDMA & OFDMA, SC-FDMA & OFDMA & GFSK & OFDMA\\
        \hline
        Bandwidth (MHz) & 1.4 -- 20 & 5 -- 100 & 1.4 -- 20 & 5 -- 100 & 20 -- 80 & 2 & 0.18 \\
        \hline
        Subcarrier Spacing & 15 kHz & 15, 30 kHz & 15 kHz & 15, 30 kHz & 78.125 kHz & N/A & 15 kHz\\
        \hline
        Modulation & QPSK--64QAM & QPSK--256QAM & QPSK--64QAM & BPSK--64QAM & BPSK--1024QAM & GFSK & QPSK\\        
        \hline
        Duplex Mode & FDD & FDD & FDD & FDD & TDD & TDD & FDD\\
        \hline
        Devices & 1 & 1 & 1-8 & 1-14 & 1-9 & 1-14 & 1\\
        \hline
    \end{tabular}
\end{center}
\end{table*}

\subsubsection{Narrowband Internet-of-Things (NB-IoT)}
NB-IoT downlink signals were generated to be release 13 compliant using the MATLAB LTE Toolbox \cite{matlabLTE}. The NB-IoT signals were generated in stand-alone mode, as opposed to the guard-band or in-band modes.

\subsection{Unknown Signal Dataset}\label{sec:unknown_dataset}
This dataset is made up of 480 10 ms signals, matching the 96 test signals per class contained in the known signal dataset.
In similar fashion to the known signal dataset, each signal is bootstrap sampled 30 times for the CNN classifier, resulting in 14400 signals total.
\subsubsection{OFDM, SC-FDMA, \& SC}
These waveforms are used in many of the aforementioned standards in the dataset of known signals. However, the signals used in the outlier set were generic, and lacked the synchronization and control signals, as well as any higher level structures found in the standards. Signal modulation was varied between QPSK, 16PSK, 64PSK, 16QAM, 64QAM, and 256QAM and signal bandwidth was varied between 25-100 MHz. The subcarrier spacings of the OFDM and SC-FDMA signals were 15, 30, or 60 kHz. This means the OFDM and SC-FDMA signals could be considered a worst-case scenario for unknown classes because they use the same modulations, subcarrier spacings, and bandwidths as some LTE and 5G signals.

\subsubsection{AM \& FM}
The AM and FM signals were the only analog modulations used in the known and unknown classes. Single and double side-band AM signals were generated.

\subsection{Over-The-Air Dataset}
A separate over-the-air (OTA) dataset was created to test the performance of the classifiers outside of a simulated environment. An NI PXIe-5645R Vector Signal Transceiver (VST) was used to transmit a 10 ms signal from each wireless standard on repeat at a carrier frequency of 3 GHz. Then, a Zynq UltraScale+ RFSoC ZCU208 development board was used to capture the signals in the 3 GHz band at a sampling rate of 125 MHz. The transmit power and receive gain were tuned such that the received signals had approximately an in-band SNR of 10 dB. A 134 ms duration capture was taken of each wireless standard. Then, bootstrap sampling was used to create 1000 1 ms signals from each capture, totalling in 7000 signals. 

\subsection{Dataset Augmentation}
Data augmentation is used to improve the generalizability of deep learning classifiers. Wireless channel and receiver models were applied to the known and unknown datasets introduced in sections \ref{sec:known_dataset} and \ref{sec:unknown_dataset} to simulate over-the-air signals.

\subsubsection{Time Delay} 
The simulated signals should have a random delay due to a lack of synchronization between the transmitter and receiver. This is achieved by using bootstrap sampling to randomly choose blocks of contiguous I/Q samples from each signal. This approach outputs signals with randomized start times and a sample length that matches the input of the CNN.

\subsubsection{Multipath Fading Channel} 
Multipath fading channels model the time-varying responses of communication channels and the effects of multiple signal paths with different phases and delays interfering at the receiver \cite{proakisDigiComm}. Ricean and Rayleigh channels are used to model line-of-sight (LOS) and non-LOS multipath fading channels respectively. The channel for each signal is randomly chosen with equal likelihood to be Rayleigh or Ricean and frequency-flat or frequency-selective. For Ricean channels, the $K$ factor is chosen randomly between 1 and 10. All channels use the ``Jakes" doppler spectrum model with a maximum doppler spread chosen randomly between 50 and 200 Hz. For frequency-selective channels, the number of discrete paths is randomly chosen between 2 and 22, the path delays are randomly chosen with a maximum delay of 200 ns, and the path gains decay randomly with a minimum path gain of -20 dB. 

The synthetic signal datasets used for tests involving the expert feature and hybrid classifiers use the 5G tapped delay line fading models A through E \cite{nrTDLfading} instead of the randomized Ricean and Rayleigh channels to model typical fading severity.

\subsubsection{In-band AWGN Channel} 
AWGN channels model noise present in communication channels as a zero-mean Gaussian random process \cite{proakisDigiComm}. 
The signal to noise ratio (SNR) in decibels (dB) is used to measure the amount of noise added to a signal.
However, since all signals are sampled at 125 MHz, some of the AWGN will be out-of-band of the signal of interest. 
In this situation, the SNR metric becomes misleading because narrowband signals will contain significantly less in-band noise compared to a wideband signal at the same SNR. In the most extreme case, an NB-IoT signal would have approximately 27 dB less in-band noise power compared to a 100 MHz 5G signal at the same SNR.

To address this, we define \textit{in-band} SNR to be the signal to noise ratio inside the 99\% power bandwidth, $B$, of the signal. Given a sampling rate, $F_s$, that oversamples the signal of interest, the relationship between in-band SNR and SNR in linear units is:
\begin{equation}
    \text{SNR}_{\text{in-band}} = \frac{F_s}{B} \, \text{SNR}
\end{equation}
Therefore, in-band SNR is equivalent to SNR after filtering out-of-band noise, making it a more accurate metric for oversampled signals.
For CNN training, signals are randomly impaired with AWGN between -5 to 20 dB in-band SNR. For testing, -20 to 20 dB in-band SNR is considered.

\subsubsection{I/Q Power Imbalance} 
I/Q imbalances are added to each signal to model mismatches in the In-phase (I) and Quadrature (Q) signal processing circuits of a radio receiver.
I/Q power imbalances are measured as the ratio in dB between the power of the I and Q samples. The I/Q imbalance of each signal is randomly chosen between -3 and 3 dB.

\subsubsection{Spurious and DC Components}
Radio front-ends can produce spurious and DC signal components caused by amplifier intermodulation and DC offsets introduced by the circuitry. The spurious signal components are modeled as a complex sinusoidal tone with random frequency and random power between -3 and 3 dB of the maximum power frequency bin in the signal of interest.



\section{Results}
In this section, the performance of the proposed CNN classifier, expert feature classifiers, and hybrid classifier are examined.
First, the CNN classifier is tested to determine the impact of frequency isolation, spatial isolation of co-frequency signals, open set classification, and captured OTA signals on classification accuracy.
Next, the accuracy of the expert feature classifiers and the impact of signal detection and frequency isolation is explored. 
Finally, the accuracy of the proposed hybrid classifier is compared against the deep learning classifier and expert feature classifiers.
In all tests, missed signal detections are scored as incorrect, to characterize the performance of the entire classification system.

\subsection{Impact of Signal Detection and Frequency Isolation on Closed Set CNN}
In this test, the impact of signal detection and frequency isolation is explored by training two CNN models on two variations of the known signal dataset introduced in section \ref{sec:known_dataset}.
In Dataset A, the signals do not undergo signal detection and frequency isolation.
Dataset B uses the same signals as Dataset A, but signal detection and frequency isolation are applied.
CNN model A is trained on Dataset A and CNN model B is trained on Dataset B. 
Dataset B models reality better because signal detection and frequency isolation are necessary to find signals of interest and isolate them from adjacent frequency bands. 

\begin{figure}[t]
    \centering
    \includegraphics[width=\linewidth]{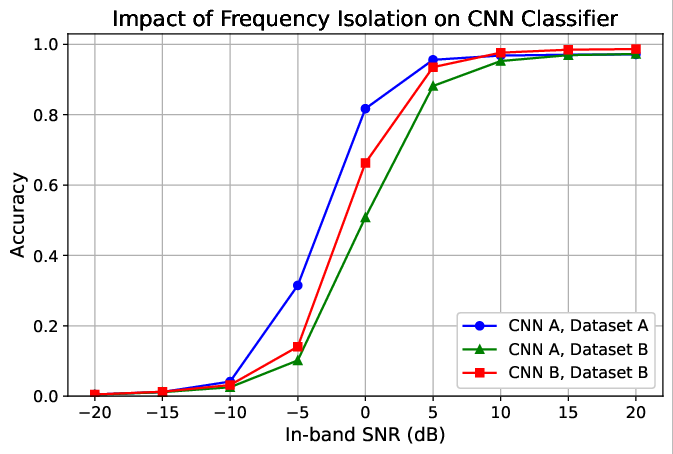}
    \caption{Closed set accuracy of CNN A and B on test signals from Datasets A and B. CNN A is trained on Dataset A, which does not use signal detection or frequency isolation. CNN B is trained on Dataset B, which uses signal detection and frequency isolation.}
    \label{fig:fssAccPlot}
\end{figure}

Figure \ref{fig:fssAccPlot} shows the accuracy of CNN A on the test data from Dataset A and Dataset B, and CNN B on the test data from Dataset B. CNN A tested on Dataset A has the highest accuracy in most cases. This is because the signals in Dataset A are under ideal circumstances, and are not impacted by error incurred by estimating the lowpass filter cutoff frequency or center frequency. When CNN A is tested with the more realistic Dataset B, the accuracy in terms of the SNR is up to 5 dB worse than Dataset A. Above 15 dB SNR, the accuracy of CNN A on Dataset A and B both converge to around 97\% accuracy. This is because the error of bandwidth and center frequency estimation are minimal at high SNR.

Additionally, Figure \ref{fig:fssAccPlot} demonstrates that some of the accuracy lost through realistic signal detection and isolation can be gained back through training dataset augmentation. CNN B is trained on signals that are isolated using the same signal detection and frequency isolation algorithms used for the test signals in Dataset B. This results in up to 4 dB SNR improvement compared to CNN A tested on Dataset B, but is still worse than the ideal case. This shows that CNN algorithm is unable to learn to mitigate all of the distortion added to the signals. In particular, at low SNR with frequency selective fading, the energy detector does a poor job of estimating the signals center frequency and bandwidth, causing a large drop in accuracy. At high SNR, CNN B slightly outperforms CNN A, approaching 98.5\% accuracy. This may be caused by frequency isolation influencing the features learned by the CNN, but the difference is small enough that it could be attributed random variations during training (i.e. different weight initialization). All other results shown in this work use CNN B as the CNN classifier.

\begin{figure}[t]
        \centering
        \includegraphics[width=0.8\linewidth]{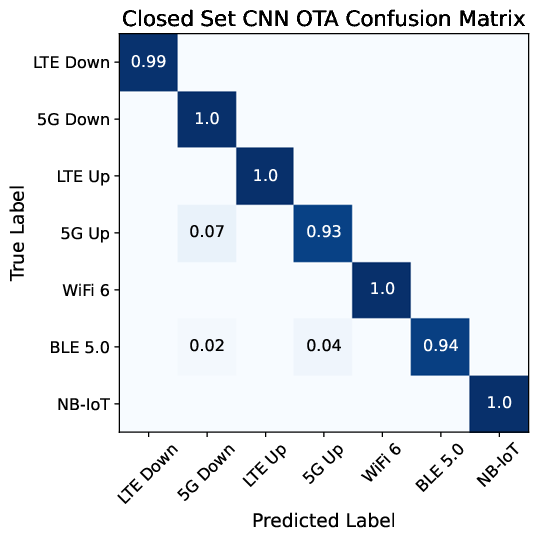}
        \caption{Closed set confusion matrix of CNN classifier on OTA dataset captured at approximately 10 dB in-band SNR. OTA signals are automatically detected and frequency isolated prior to classification.}
        \label{fig:ClosedSetConfu_ota}
\end{figure}

The accuracy of CNN B can be verified using the OTA dataset. Figure \ref{fig:ClosedSetConfu_ota} shows the closed set accuracy of the CNN tested with OTA data via a confusion matrix. The average accuracy on the OTA dataset is 98\%, which aligns with the 97.6\% average accuracy on the synthetic dataset at 10 dB SNR. This confirms that the signal and receiver modeling accurately models the distortion caused by hardware imperfections and the wireless channel.

\subsection{Co-Frequency Signal Classification with Closed Set CNN}

\begin{figure}[t]
    \centering
    \includegraphics[width=\linewidth]{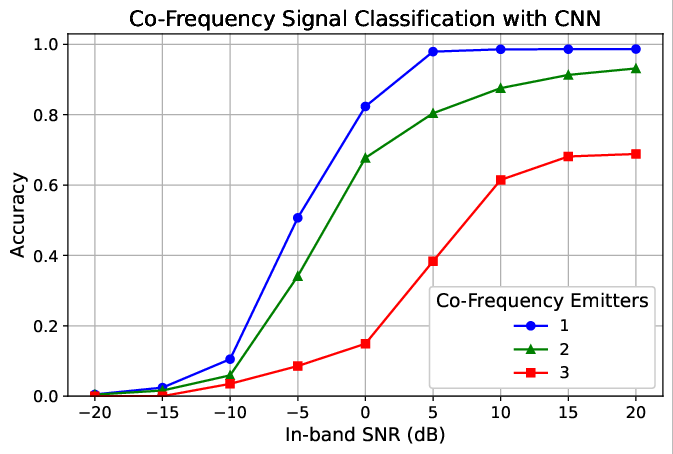}
    \caption{Closed set CNN classification accuracy of co-frequency emitters using model order estimation, direction of arrival estimation, and spatial filtering with a four element ULA.}
    \label{fig:mmseMonteCarlo_k1_k2_k3}
\end{figure}

In this test, co-frequency signals are simulated from random spatial directions between -80 and 80 degrees with respect to the boresight of a four element ULA. Results for one, two, and three co-frequency emitters are shown in Figure \ref{fig:mmseMonteCarlo_k1_k2_k3}. The co-frequency signals are randomly combined from the test signals in the known dataset introduced in section \ref{sec:known_dataset}, such that all combinations of signals classes are tested. 
The power per Hz of the co-frequency signals are normalized to model realistic signal powers. 
For example, a 180 kHz bandwidth NB-IoT signal would have 20 dB lower power than a 20 MHz bandwidth 5G signal, due to the bandwidth being approximately a 100 times smaller. The co-frequency signals are spatially isolated using model order estimation (MOE), direction of arrival estimation (DOA), and spatial filtering. After spatial signal isolation, all detected signals are independently frequency isolated and classified by the CNN. 

In the case of a single emitter in Figure \ref{fig:mmseMonteCarlo_k1_k2_k3}, the use of spatial isolation increases accuracy at low SNR compared to the single antenna accuracy of CNN B shown in Figure \ref{fig:fssAccPlot}. Theoretically, the MMSE beamformer with a four element ULA yields a 6 dB SNR gain compared to single antenna. However, in practice, the necessity of MOE and DOA estimators reduce the SNR gain to 3-5 dB on average. In the case of two co-frequency emitters, the single antenna CNN would be incapable of correctly classifying both signals correctly due to the interference.
However, after spatial signal isolation, the CNN is able to classify the interfering signals with 93\% accuracy at 20 dB in-band SNR.  
When there are three co-frequency emitters, the accuracy declines significantly to 68\% at 20 dB in-band SNR.
This is partially because adding more interfering signals increases the likelihood that signals have close angular spacing and makes weak signals more difficult to detect.
In a previous work \cite{milcomPaper2}, we showed that using a sparse linear array increases classification accuracy in scenarios with many emitters, close angular spacing between emitters, or large power imbalances without increasing the number of array elements.

\subsection{Open Set CNN}
\begin{figure}[t]
        \centering
        \includegraphics[width=\linewidth]{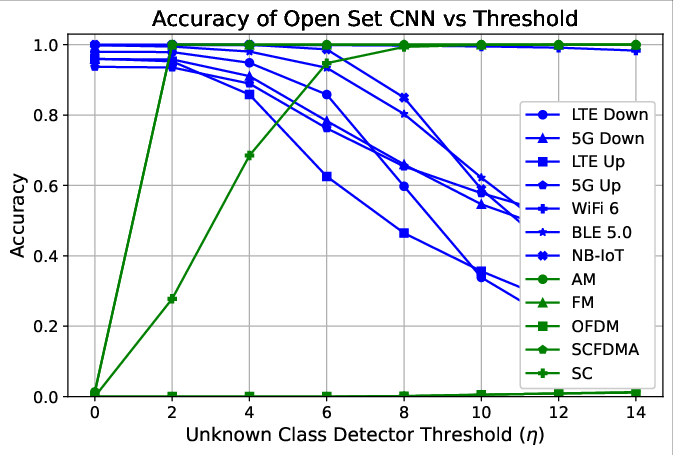}
        \caption{Accuracy of open set classifier over different unknown class detector thresholds at an in-band SNR of 10 dB. Blue signal classes are known (represented in the training data) and green signal classes are unknown. Unknown classes are correctly classified when detected as unknown.}
        \label{fig:OpenAccVsThresh}
\end{figure}

In the open set case, signals from unknown classes must be identified as unknown to appropriately categorize the signal environment. 
The performance of the unknown class detector over different threshold values is characterized in Figure \ref{fig:OpenAccVsThresh}. 
Increasing the unknown class detector threshold is equivalent to shrinking the decision region around each of the classes. 
Increasing the threshold results in monotonically decreasing accuracy for the known classes and monotonically increasing accuracy detecting unknown classes. 
This is because shrinking a decision regions will always result in the same or more signals being identified as unknown.

In Figure \ref{fig:OpenAccVsThresh}, both AM and FM signals are accurately recognized as unknown, even at low thresholds. 
This suggests that CNN learned features that can easily distinguish between AM and FM and the known classes.
However, AM and FM are the most different from known classes of the signal types in the unknown datasets, due to the use of analog modulation and different bandwidths.
SC, OFDM, and SCFDMA share similarities with the known standards, and therefore have lower detection rates.
SC signals can be identified as unknown with around 95\% accuracy at a threshold of 6, but OFDM and SCFDMA signals are rarely detected.
The open set classifier relies on the CNN learning a feature space that separates unknown classes from known classes.
However, good separability can be hard to achieve because the CNN cannot be directly trained on unknown classes.
In our CNN, this manifests as the inability to differentiate OFDM and SCFDMA from the known signal classes.
In previous work \cite{milcomCofPaper}, a different version of the open set CNN was able to detect OFDM and SCFDMA signals as unknown. However, it was found that after signal detection and isolation the CNN could no longer reliably identify OFDM or SCFDMA signals.
We will show later that expert feature classifiers can provide more robust unknown class detection and explainability for missed detections.

One method for choosing the unknown class detector threshold is to pick an accuracy target for known classes and increase the threshold until the accuracy meets the target. 
We arbitrarily choose an average accuracy target of 85\%, resulting in a maximum threshold of 6.
At this threshold with 10 dB in-band SNR, AM and FM signals are detected with 100\% accuracy, SC signals are detected with 95\% accuracy, and OFDM and SCFDMA are detected at a rate of less than 1\%.

\begin{figure}[t]
    \centering
        \includegraphics[width=0.8\linewidth]{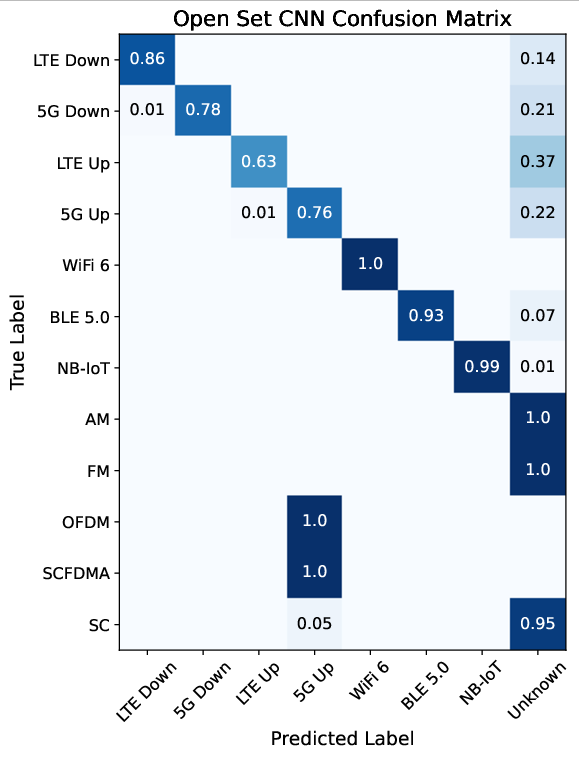}
        \caption{Open set CNN confusion matrix with unknown class detector threshold $\eta$ = 6 and in-band SNR = 10 dB}
        \label{fig:OpenSetConfu}
\end{figure}

\begin{figure}[t]
        \centering
        \includegraphics[width=0.8\linewidth]{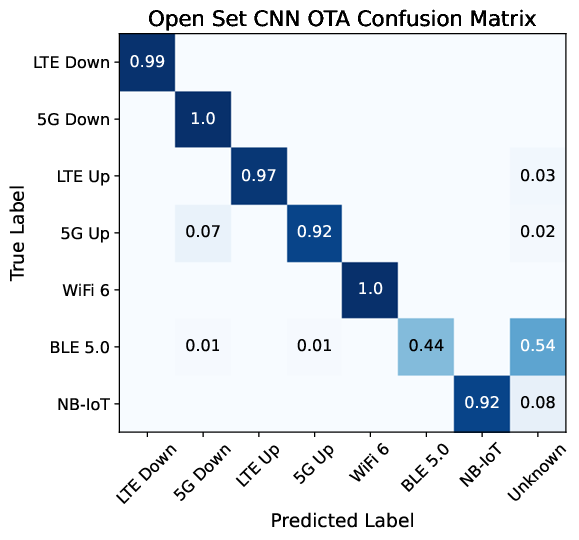}
        \caption{Open set CNN confusion matrix on OTA dataset captured at approximately 10 dB in-band SNR with unknown class detector threshold $\eta$ = 6.}
        \label{fig:OpenSetConfu_ota}
\end{figure}

Confusion matrices of the open set CNN with a threshold of 6 are shown in Figures \ref{fig:OpenSetConfu} and \ref{fig:OpenSetConfu_ota} for the synthetic datasets and the OTA dataset respectively. 
The accuracy of the open set CNN is reduced for known classes because known signals are mistakenly detected as unknown. 
For the synthetic test data in Figure \ref{fig:OpenSetConfu}, LTE and 5G signals are most commonly erroneously detected as unknown. 
OFDM and SCFDMA are consistently classified as 5G uplink, meaning that these signals are similar in the learned feature space of the CNN.
This could be because the OFDM and SCFDMA signals share common modulations and bandwidths as the 5G uplink signals. 
For the OTA data in Figure \ref{fig:OpenSetConfu_ota}, LTE and 5G signals are detected as unknown with a much lower rate than for the synthetic data.
This could be because the OTA dataset used a narrower set of signal parameters than the synthetic datasets, suggesting that some combinations of signal parameters are more likely to be detected as unknown than others.

\begin{figure}
    \centering
    \includegraphics[scale=0.49]{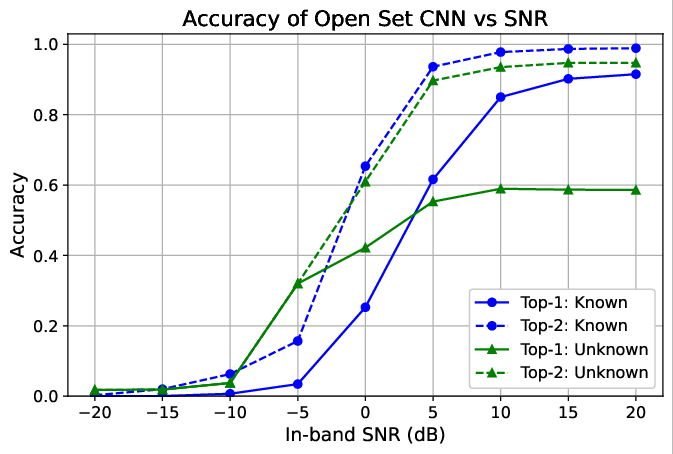}
    \caption{Average accuracy of open set classifier with unknown class detector threshold $\eta$ = 6. Top $M$ accuracy counts the classification as correct if the true class is in the top $M$ classes predicted by the classifier. The blue curves are for known classes and the green curves are for unknown classes.}
    \label{fig:openAccTop2}
\end{figure}

The top-1 and top-2 accuracy of the open set CNN are shown in Figure \ref{fig:openAccTop2}. 
Up to this point, all of the results shown have been top-1 accuracy, meaning the true class needed to be the class predicted with the highest probability to be scored as correct. 
The cost of using the open set CNN is significant as the top-1 accuracy for known signals has a SNR loss over 5 dB and the maximum accuracy is decreased by 7\% compared to the top-1 closed set accuracy of CNN B in Figure \ref{fig:fssAccPlot}.
Additionally, the accuracy of detecting unknown classes is poor due to the inability to detect generic OFDM or SCFDMA signals. 
The unreliable detection of unknown signal classes motivates the augmentation of the CNN with robust expert feature classifiers.

In the hybrid classifier, the CNN is used to reduce the computational complexity of the expert feature classifiers. This is achieved by computing the most likely expert feature classifiers using the CNN's top-2 predicted classes. 
The top-2 accuracy in Figure \ref{fig:openAccTop2} shows that for known classes, the two most likely classes predicted by the CNN reaches 99\% accuracy at high SNR. For unknown classes, the top-2 accuracy of the CNN increases to 95\% at high SNR. 
Therefore, the hybrid classifier can maintain good accuracy while only running the most probable expert feature classifiers.

\subsection{Expert Feature Classifiers}
\begin{figure}[t]
    \centering
        \includegraphics[width=\linewidth]{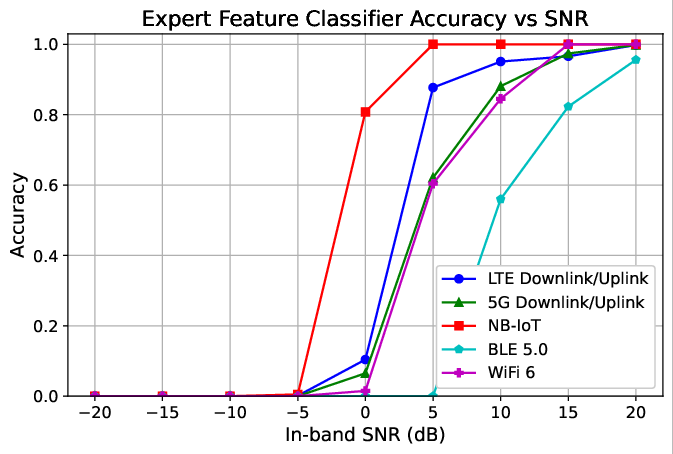}
        \caption{Classification accuracy of detection for expert feature classifiers with downconversion using an estimate of the center frequency.}
        \label{fig:featureAcc_wFSS}
\end{figure}

The expert feature classifiers are open set by definition and the missed detection rate for unknown classes is equivalent to the false alarm rates in Table \ref{tab:crcPfa}. The probabilities of detection for each of the expert feature classifiers are shown in Figure \ref{fig:featureAcc_wFSS}.
The BLE 5.0 classifier has the lowest accuracy because the transmissions do not have error correction coding. Therefore, even at moderate SNR, bit errors caused by AWGN can cause the CRC to fail.
Ultimately, all expert feature classifiers approach 100\% accuracy as the in-band SNR exceeds 20 dB and approach 0\% accuracy below -5 dB SNR.

\begin{figure}[t]
    \centering
    \includegraphics[width=\linewidth]{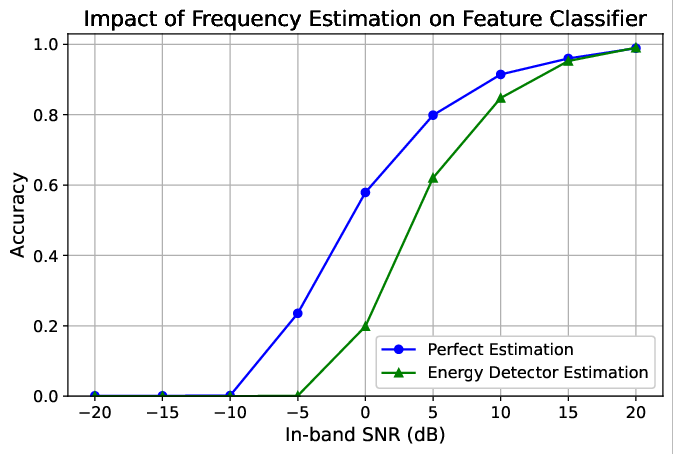}
    \caption{Average probability of detection for expert feature classifiers with perfect downconversion vs downconversion using an estimate of the center frequency from an energy detector. }
    \label{fig:avgfeatureAcc}
\end{figure}

The remaining expert feature classifiers perform similarly, but the standards with wider bandwidth signals, like LTE, 5G, and WiFi 6, have lower accuracy than NB-IoT. This is because frequency selective fading causes power variations in the spectrum of the signal, increasing the center frequency estimation error during signal detection. 
Especially at low SNR, frequency selective fading can cause portions of the signal spectrum to be below the noise floor, while other portions are detectable.
This can cause significant center frequency estimation errors that exceed the maximum correctable offset of the frequency estimation algorithms used in the expert feature classifiers.
Figure \ref{fig:avgfeatureAcc} shows that center frequency estimation error during signal detection results in an SNR loss of 5 dB below 10 dB SNR. 
To reduce the loss in accuracy, the maximum correctable frequency offset could be increased at the cost of computational complexity.
This could be accomplished by increasing the range of frequency shifts during the coarse frequency correction for standards like LTE.

\begin{table}[t]
    \begin{center}
    \caption{Computational Complexity Comparison}
    \label{tab:compuCompl}
    \footnotesize
    \begin{tabular}{|c| c | c|} 
        \multicolumn{3}{c}{\textbf{Deep Learning}}\\
        \hline
        Classifier & Additions $\times10^6$ & Multiplications $\times10^6$\\ [0.5ex] 
        \hline\hline
        CNN & 36.6 & 44.7\\
        \hline
        \multicolumn{3}{c}{\textbf{Expert Feature}}\\
        \hline
        Classifier & Additions $\times10^6$ & Multiplications $\times10^6$\\ [0.5ex] 
        \hline\hline
        4G Downlink/Uplink & 1543 & 1551\\
        \hline
        NB-IoT Downlink & 26950 & 26960\\
        \hline
        5G Downlink/Uplink & 12470 & 12510\\
        \hline
        BLE Adv. Packet & 116.3 & 114.6\\
        \hline
        WiFi 6 & 150.1 & 175.9\\
        \hline
        \hline
        \textbf{Average Known} & 31570 & 31640\\
        \hline
        \textbf{Average Unknown} & 55250 & 55370\\
        \hline
        \multicolumn{3}{c}{\textbf{Average Hybrid}}\\
        \hline
        & Additions $\times10^6$ & Multiplications $\times10^6$\\ [0.5ex]
        \hline \hline
        Minimum & 7929 & 7955\\
        \hline
        Maximum & 15820 & 15860\\
        \hline
        \end{tabular}
    \end{center}
\end{table}

\subsection{Hybrid Classifier}
We propose the hybrid classifier to combine the strengths of the CNN classifier and the expert feature classifiers.
The open set CNN achieves a high top-2 accuracy, but is unreliable for unknown class detection.
The expert feature classifiers have good accuracy for known signals and theoretically justified unknown class detection, based on the CRC probability of false alarm.
However, the expert feature classifiers require significantly higher computational complexity compared to the CNN.

Table \ref{tab:compuCompl} shows the computational complexity of each of the classifiers. The CNN, including feature preprocessing, has lower complexity than any expert feature classifier.
Compared to the CNN, the average complexity of the expert feature classifiers for known classes is over 700 times higher and the average complexity for unknown classes is over 1300 times higher than the CNN. 
The average complexity for known classes assumes that the signal of interest is equally likely to be any of the standards, so on average four expert feature classifiers must be run before a detection.
The average complexity of unknown classes assumes that all of the expert feature classifiers must be run.

The hybrid classifier leverages the low computational complexity of the CNN to find the highest probability wireless standards.
Then, only one or two expert feature classifiers, corresponding to the most likely wireless standards, are run.
The minimum hybrid complexity refers to the case when one of the top-2 CNN predictions is unknown, causing only one expert feature classifier to be run. The maximum complexity is when two expert feature classifier must be run. On average, the hybrid classifiers is at least two times lower complexity and at best nearly seven times lower complexity.

\begin{figure}[t]
        \centering
        \includegraphics[width=\linewidth]{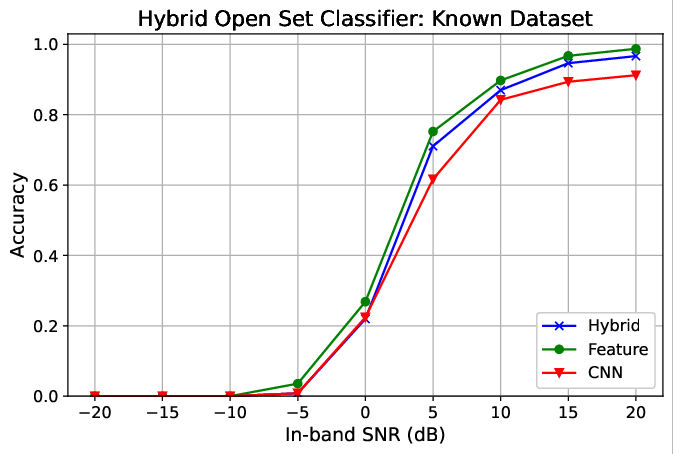}
        \caption{Classification accuracy of the hybrid, CNN, and expert feature open set classifiers on the known dataset.}
        \label{fig:hybridIn}
\end{figure}

\begin{figure}[t]
        \centering
        \includegraphics[width=\linewidth]{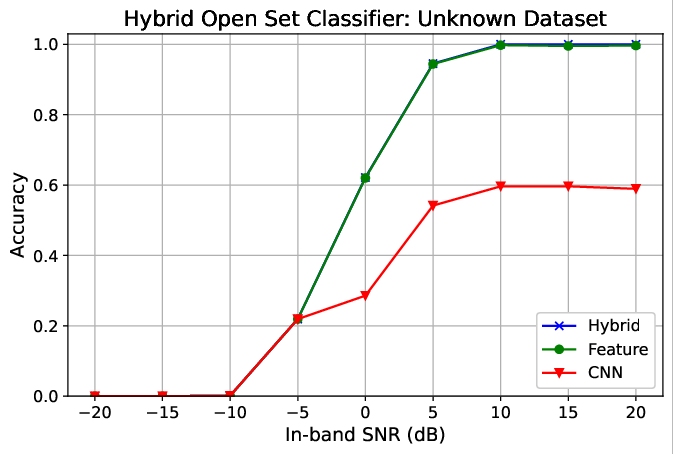}
        \caption{Classification accuracy of the hybrid, CNN, and expert feature open set classifiers on the unknown dataset.}
        \label{fig:hybridOut}
\end{figure}

The accuracy of the hybrid classifier is compared to the CNN and expert feature classifiers in Figures \ref{fig:hybridIn} and \ref{fig:hybridOut}. 
Figure \ref{fig:hybridIn} shows that the accuracy for known signal classes follows a similar trend for all classifiers. 
The expert feature classifiers have the highest accuracy, with the hybrid classifier trailing by 1-2\% above 10 dB in-band SNR. 
The accuracy is lower for the hybrid classifier because occasionally the true wireless standard is not in the top-2 inference of the CNN.
The open set CNN is approximately 3-10\% lower accuracy than the hybrid classifier above 0 dB.
For unknown signals, shown in Figure \ref{fig:hybridOut}, the CNN has significantly lower accuracy due to its inability to detect generic OFDM or SCFDMA signals as unknown. 
The hybrid and expert feature classifiers achieve close to 100\% accuracy at high in-band SNR due to the robustness of the CRC checks. 
The hybrid classifier slightly outperforms the expert feature classifiers because it requires less CRC checks per classification, resulting in a lower $P_{fa}$.
At low in-band SNR, all classifiers have low accuracy due to the energy detector failing to detect the signals.

A key consideration is whether to use the closed set or open set CNN in the hybrid classifier. The closed set CNN has higher accuracy than the open set CNN for known classes, which would reduce the loss in accuracy of the hybrid classifier compared to the expert feature classifiers. The trade-off is that the computational complexity of the hybrid classifier would increase due to the closed set CNN always predicting two wireless standards in the top-2 inference. The open set CNN has the option to classify the signal as unknown in the top-2 inference, resulting in less expert feature classifiers run when the CNN determines the probability of an unknown signal is high. We choose to use the open set CNN in the hybrid classifier to reduce computational complexity, but the best option would depend on the application.


\section{Conclusion}
In this paper, we proposed a hybrid classifier that intelligently selects expert feature classifiers using the top-2 most probable classes predicted by a CNN classifier. 
The hybrid classifier achieves 97\% accuracy for known classes and nearly 100\% accuracy of unknown classes at 20 dB in-band SNR.
Compared to the the CNN classifier, the hybrid classifier has 5\% higher accuracy for known classes and 40\% higher accuracy for unknown classes at 20 dB SNR.
The drastic improvement in unknown class detection is due to the reliability of the CRC checks in the expert feature classifiers.
Compared to the expert feature classifier, the hybrid classifier has similar accuracy for both known and unknown classes, but uses two to seven times lower computational complexity.
The hybrid classifier achieves this by synergistically combining the strengths of each classifier architecture.

We fully model signal detection and isolation in time, frequency, and space. Automatic frequency isolation of signals resulted in roughly a 5 dB SNR loss for the CNN and expert feature classifiers. However, the CNN could reclaim up to 4 dB of the SNR loss through dataset augmentation.
Spatial signal isolation gives the classifiers the ability to classify co-frequency signals from multiple emitters. 
With two co-frequency emitters, the interfering signals can be detected, isolated, and classified with 93\% accuracy using the closed set CNN.
The addition of realistic signal detection and isolation and robust open set classification makes the hybrid classifier promising for real-world wideband spectrum sensing.

An area of open research is developing better methods of understanding the false alarm rate for open set deep learning models. 
Current approaches rely on an empirical analysis to characterize the performance, but this depends on the quantity and quality of the data.
The ability to guide the feature learning in deep learning models with theory would lead to better insights on the feature space of the classifiers, and the resulting open set performance.


%
\bibliographystyle{ieeetr}
\bibliography{ref.bib}

\vfill

\end{document}